\documentclass[aps, pra, twocolumn, 10pt, showpacs, floatfix, longbibliography]{revtex4-1}
\usepackage{graphicx, verbatim, amsmath, amssymb,color,setspace,subfigure,braket,url}

\definecolor{darkblue}{rgb}{0.1,0.2,0.6} \definecolor{darkred}{rgb}{0.8,0.1,0.2} \definecolor{darkorange}{rgb}{0.7,0.4,0.1}
\usepackage[colorlinks,citecolor=darkblue,linkcolor=darkred,urlcolor=darkblue]{hyperref}

\usepackage{soul}

\begin{document}

\title{Expansion of one-dimensional lattice hard-core bosons at finite temperature}

\author{Wei Xu}
\affiliation{Department of Physics, Pennsylvania State University, University Park, Pennsylvania 16802, USA}
\author{Marcos Rigol}
\affiliation{Department of Physics, Pennsylvania State University, University Park, Pennsylvania 16802, USA}

\begin{abstract}
We develop an exact approach to study the quench dynamics of hard-core bosons initially in thermal equilibrium in one-dimensional lattices. This approach is used to study the sudden expansion of thermal states after confining potentials are switched off. We find that a dynamical fermionization of the momentum distribution occurs at all temperatures. This phenomenon is studied for low initial site occupations, for which the expansion of the cloud is self-similar. In this regime, the occupation of the natural orbitals allows one to distinguish hard-core bosons from noninteracting fermions. We also study the free expansion of initial Mott insulating domains at finite temperature, and show that the emergence of off-diagonal one-body correlations is suppressed gradually with increasing temperature. Surprisingly, the melting of the Mott domain is accompanied by an effective cooling of the system. We explain this phenomenon analytically using an equilibrium description based on an emergent local Hamiltonian.
\end{abstract}

\pacs{67.85.-d, 67.85.De, 03.75.Kk}

\maketitle

\section{Introduction}
In recent years, experimental and theoretical studies of far-from-equilibrium dynamics of ultracold atoms have led to the discovery and understanding of a large number of unconventional nonequilibrium effects, among which many occur during expansion dynamics in the presence of strong interactions. This is usually achieved by turning off confining potentials and letting atoms expand in optical lattices~\cite{Schneider_12, Ronzheimer_13, Xia_15, Vidmar_Ronzheimer_15}. Of direct relevance to this work, theoretical studies of the expansion of hard-core bosons in one dimension (1D) predicted the occurrence of a dynamical fermionization of the momentum distribution~\cite{Rigol_Muramatsu_05b, Minguzzi_05, Rigol_Muramatsu_05c} and of quasicondensation at finite momenta when the expansion begins from Mott insulating domains at unit filling~\cite{Rigol_Muramatsu_04b, Rigol_Muramatsu_05c, Micheli_04, Micheli_06}. The latter was recently observed experimentally~\cite{Vidmar_Ronzheimer_15}. The effect of finite on-site (or nearest neighbor) interactions has been explored in bosonic (or spin) and fermionic systems in Refs.~\cite{gobert05, rodriguez_manmana_06, Meisner_08, Langer_12, bolech_meisner_12, Vidmar_13, sabetta_misguich_13, mei_vidmar_16, bertini_collura_16, deluca_collura_16}, while changes induced by increasing dimensionality have been explored in Refs.~\cite{Hen_10, Jreissaty_11, Vidmar_13, Hauschild_15}. Another topic that has attracted much attention is the effect of multiple occupancies in the expansion in the presence of strong interactions \cite{Meisner_09, Muth_12, Jreissaty_13, Vidmar_13, Boschi_14, Andraschko_15}.
 
A common theme in the theoretical studies mentioned above is that the expansion always starts from initial pure states. It is in general unknown whether dynamical phenomena observed for pure states are robust enough to survive in experimentally realistic situations, in which initial states are expected to be in thermal equilibrium at nonzero (and sometimes not low) temperatures. Recently, a finite-temperature hydrodynamic approach~\cite{Bouchoule_16b} and an exact Fredholm determinant approach~\cite{Atas_16a,Atas_16b} were developed to study breathing-mode oscillations and collective many-body bounces. The theoretical calculations were found to agree well with experimental results in Ref.~\cite{Fang_14}.

In this work, we explore finite-temperature effects in the expansion of hard-core bosons in 1D lattices. We introduce an exact approach to study dynamics based on an existing finite-temperature equilibrium method~\cite{Rigol_05}. We find that the dynamical fermionization of the momentum distribution of hard-core bosons, originally discussed in the expansion from the ground state~\cite{Rigol_Muramatsu_05b, Minguzzi_05, Rigol_Muramatsu_05c}, occurs at all temperatures. For this effect, we focus on initial states with low site occupations, for which the cloud expands in a self-similar way. In contrast to the momentum distribution, the natural orbital occupations $\lambda(\eta)$ remain unchanged during the expansion and display the universal $\eta^{-4}$ tail known from equilibrium calculations \cite{Rigol_Muramatsu_04a}. This means that, during the expansion, the natural orbital occupations allow one to distinguish hard-core bosons from noninteracting fermions. 

We also study the expansion from initial Mott insulating domains. We show that while one-body correlations are enhanced during the dynamics, they are suppressed as the initial temperature is increased. The enhancement of such correlations is reminiscent of the dynamical quasicondensation shown to occur in pure states \cite{Rigol_Muramatsu_04b}. Nonzero temperatures have an effect in the expanding systems that is similar to that in systems in equilibrium~\cite{Rigol_05}, namely, power-law correlations are replaced by exponential ones. Surprisingly, we find that those systems appear to cool down during the melting of the Mott domains. They are effectively in local thermal equilibrium with temperatures that decrease with time. We explain this analytically using an emergent local Hamiltonian~\cite{Vidmar_Iyer_15}. The cooling can be intuitively understood to be the result of energy being transferred from internal energy to center-of-mass motion. We construct the corresponding reference Hamiltonian~\cite{Meisner_08, Eisler_09} in order to illustrate this picture.

The presentation is organized as follows. In Sec.~\ref{Sec:II}, we introduce the numerical approach developed to study dynamics of finite-temperature initial states. We study the effect of temperature in the fermionization of the momentum distribution during the expansion in Sec.~\ref{Sec:III}. Section~\ref{Sec:IV} is devoted to investigate the melting of an initial Mott insulating domain created in thermal equilibrium in the presence of a linear potential. We present a detailed discussion of the emergent local Hamiltonian solution in Sec.~\ref{Sec:IV}(B), and of the intuitive understanding of the effective cooling in terms of a reference Hamiltonian in Sec.~\ref{Sec:IV}(C). Finally, a summary is presented in Sec.~\ref{Sec:V}.

\section{Model and Method}\label{Sec:II}

Hard-core bosons provide an effective description of the Bose-Hubbard model in the limit of infinite on-site repulsion \cite{cazalilla_citro_review_11}. The hard-core boson Hamiltonian reads
\begin{eqnarray}\label{eq:ham}
 \hat{H}&=&\hat{H}_0+\hat{H}_V \\
 \hat H_0&=&-t\sum_l(\hat b_{l+1}^\dag \hat b_l+\text{H.c.}),\  \text{and}\  \hat H_V=V_\alpha\sum_ll^\alpha \hat n_l \nonumber
\end{eqnarray}
where $\hat H_0$ models nearest-neighbor hoppings with amplitude $t$, and $\hat H_V$ models a confining potential. The hard-core boson creation ($\hat b_l^\dag$) and annihilation ($\hat b_l$) operators satisfy standard bosonic commutation relations with the constraint $(\hat b_l^\dag)^2=\hat b_l^2=0$. In $\hat H_V$, $\alpha$ denotes the exponent of the power-law potential, e.g., $\alpha=2$ stands for the traditional harmonic confinement. In what follows, we set $t=k_B=\hbar=1$ and the lattice spacing $a=1$. Hamiltonian \eqref{eq:ham} can be mapped onto the spin-1/2 XX model via the Holstein-Primakoff transformation and then onto non-interacting spinless fermions via the Jordan-Wigner transformation \cite{cazalilla_citro_review_11}
\begin{equation}
\hat b_l^\dag=\hat f_l^\dag\prod_{\beta=1}^{l-1}e^{-i\pi \hat f_\beta^\dag \hat f_\beta} \;,\;
\hat b_l=\prod_{\beta=1}^{l-1}e^{i\pi \hat f_\beta^\dag \hat f_\beta}\hat f_l\;.
\end{equation}
The corresponding fermionic Hamiltonian shares the same form with that of the hard-core bosons (up to a possible boundary term) with $\hat b_l^\dag$ ($\hat b_l$) replaced by $\hat f_l^\dag$ ($\hat f_l$), signaling the Bose-Fermi duality in 1D~\cite{Girardeau_60}. This means that hard-core bosons and fermion share the same thermodynamic properties for diagonal (site-occupation related) observables, while nontrivial differences appear in off-diagonal one-body observables.

To study the time evolution of finite-temperature initial states, we consider the grand-canonical ensemble, for which we devise a computational approach that has a computation time that scales polynomially with system size. To avoid the particle-number dependence of the boundary term after the mapping, we restrict our analysis to open chains~\cite{Rigol_05}. The equal-time one-body density matrix after time $\tau$ of the evolution can be written as
\begin{equation}
\begin{split}
&\rho_{ij}(\tau)=\text{Tr}\left[e^{i\hat H_F\tau}\hat b_i^\dag \hat b_je^{-i\hat H_F\tau}\hat \rho_I\right] \\
&=\text{Tr}\left[e^{i\hat H_F\tau}\prod_{\alpha=1}^{i-1}e^{-i\pi \hat n_\alpha^f}\hat f_i^\dag \hat f_j\prod_{\beta=1}^{j-1}e^{i\pi \hat n_\beta^f}e^{-i\hat H_F\tau}\hat \rho_I\right]
\end{split}
\label{eq:ETOBDM_FT}
\end{equation}
where the density matrix of the initial thermal state is $\hat \rho_I=e^{-(\hat H_I-\mu N)/T}/Z$, $Z=\text{Tr}[e^{-(\hat H_I-\mu N)/T}]$ is the partition function, $T$ is the initial temperature, and we denote $\hat H_I$ ($\hat H_F$) as the initial (final) Hamiltonian. 

A numerical evaluation of $\rho_{ij}(\tau)$ in Eq.~\eqref{eq:ETOBDM_FT} can be carried out using the fact that the trace of exponentials of bilinear forms in fermionic creation and annihilation operators over Fock space can be written as~\cite{Rigol_05}
\begin{equation}
\begin{split}
&\text{Tr}\left[\text{exp}\left(\sum_{ij}\hat f_i^\dag X_{ij}\hat f_j\right)\text{exp}\left(\sum_{kl}\hat f_k^\dag Y_{kl}\hat f_l\right)...\right] \\ 
&=\text{det}[I+e^Xe^Y...]\;.
\end{split}
\label{eq:DQMC_II}
\end{equation}

For the off-diagonal matrix elements, $\rho_{ij}$ with $i\neq j$, one can write~\cite{Rigol_05}
\begin{equation}
 \hat f_i^\dag \hat f_j=\text{exp}\left(\sum_{kl}\hat f_k^\dag A_{kl}\hat f_l\right)-1\;,
\label{eq:exp_ij}
\end{equation}
where matrix $A$ has only one nonzero element $A_{ij}=1$. Thus, substituting $\hat f_i^\dag \hat f_j$ in Eq.~\eqref{eq:ETOBDM_FT} results in the need to evaluate two determinants, one involving $e^A=I+A$ and the other one involving $I$. The strings arising from the Jordan-Wigner transformation contribute a diagonal matrix $O_i$ ($O_j$) with the first $i-1$ ($j-1$) diagonal elements equal to -1 and the rest equal to 1. 

Putting all this together, and carrying out a few further simplifications that lead to improvements in the speed and numerical stability of the computations, results in
\begin{eqnarray}\label{eq:ETOBDM_FT_ij}
\rho_{ij}(\tau)&=&\frac{(-1)^{i-j}}Z\bigg\{\text{det}\bigg[U_I^\dag e^{iH_F\tau}O_j(I-A) \\
&&\hspace{2cm}\times O_ie^{-iH_F\tau}U_I+e^{-(E_I-\mu I)/T}\bigg]\nonumber \\
&& -\text{det}\bigg[U_I^\dag e^{iH_F\tau}O_jO_ie^{-iH_F\tau}U_I+e^{-(E_I-\mu I)/T}\bigg]\bigg\}, \nonumber
\end{eqnarray}
where $H_IU_I=U_IE_I$ ($H_FU_F=U_FE_F$), with $E_I$ ($E_F$) being the diagonal matrix with the sorted eigenenergies of the initial (final) Hamiltonian. The matrix representation of the time evolution operator $e^{-i\hat H_F\tau}$ can be obtained as $e^{-iH_F\tau}=U_Fe^{-iE_F\tau}U_F^\dag$. 

The diagonal matrix elements $\rho_{ii}$ can be calculated in a similar fashion, starting from the following property of the exponential $\exp\left(-i\pi \hat f_i^\dag \hat f_i\right)=1-2\hat f_i^\dag \hat f_i$. Another, more efficient, way to obtain $\rho_{ii}$ follows from the fact that $\rho_{ii}$ is identical for non-interacting fermions and hard-core bosons. For noninteracting fermions one has
\begin{equation}
 \rho_{ii}(\tau)=1-\left[e^{iH_F\tau}\left(I+e^{-(H_I-\mu I)/T}\right)e^{-iH_F\tau}\right]_{ii}^{-1}\;.
\label{eq:OBDM_FT_ii}
\end{equation}
This is the equivalent of Eq.~(17) in Ref.~\cite{Rigol_05} after correcting a typographical error.

\section{Fermionization}\label{Sec:III}

As mentioned before, due to the Bose-Fermi mapping, diagonal observables are identical in hard-core and spinless fermion systems~\cite{Girardeau_60}. Off-diagonal ones, such as two-point one-body correlations, on the other hand, are different. This leads to starkly different behaviors of the momentum distribution of hard-core bosons and fermions. In the ground state, the former exhibit quasicondensation at zero momentum (the occupation of the zero momentum mode scales with $\sqrt{N}$, where $N$ is the number of bosons in the system)~\cite{cazalilla_citro_review_11}. This cannot occur with noninteracting fermions because of the Pauli exclusion principle. In addition, the hard-core boson momentum distribution function exhibits high-momentum tails that decay as $k^{-4}$ \cite{minguzzi_vignolo_02, olshanii_dunjko_03, Rigol_Muramatsu_04a}. Such tails are not present in the fermionic counterpart, in which the Fermi momentum cuts off the distribution. Studies of the momentum distribution of hard-core bosons during expansion dynamics have revealed that it approaches the momentum distribution of the fermionic system to which they can be mapped~\cite{Rigol_Muramatsu_05b, Minguzzi_05, Rigol_Muramatsu_05c}. This has been explained in terms of a dynamical phase that appears during the expansion, and which results in an asymptotic momentum distribution of the hard-core bosons that is identical to the momentum distribution of the fermions in the external harmonic confinement~\cite{Minguzzi_05}. Here, we show that such dynamical fermionization also occurs if the initial state is at nonzero temperature.

\subsection{Momentum distribution}

To study the dynamics, we start from a harmonically trapped system in thermal equilibrium. For sufficiently large systems, the initial state is characterized by the characteristic density $\tilde{\rho}=N/\zeta$ and the temperature $T$, where $\zeta=(V_\alpha)^{-1/\alpha}$ is the length scale introduced by the confining potential ($\alpha=2$ for the harmonic trap). We are interested in the momentum distribution during the expansion after turning off the harmonic trap, which we calculate as~\cite{Rigol_Muramatsu_05a, Rigol_Muramatsu_04a, Rigol_05}
\begin{equation}
 n_k(\tau)=\frac 1 \zeta \sum_{jl}e^{ik(x_j-x_l)}\rho_{jl}(\tau).
\end{equation}
In the thermodynamic limit and finite site occupations, a scaling solution with constant $[\tilde{\rho},\;T,\;\tau/N]$ characterizes the system~\cite{Rigol_05,Rigol_Muramatsu_05b}. In the low-site-occupation limit, the lattice system maps onto a continuous system and a scaling solution with a reduced number of parameters $[T/\tilde{\rho},\;\tau\tilde{\rho}/N]$ exists. This is discussed in Ref.~\cite{xu_rigol_15} for the equilibrium case. (Note that the characteristic density defined here is different from the one defined in Ref.~\cite{xu_rigol_15}, where the focus were systems in the continuum.)

In this section, we focus on systems with low site occupations, so that they can be well described using scaling results from systems in the continuum. A previous study for the ground state indicates this is the case for $\tilde{\rho}\lesssim0.2$~\cite{Rigol_Muramatsu_05b}. Analytic expressions for the time-dependent one-body density matrix during the expansion starting from the ground state in the continuum were derived in Ref.~\cite{Minguzzi_05} and were recently generalized to finite-temperature situations in Refs.~\cite{Atas_16a,Atas_16b}. From the scaling transformation, the equal-time one-body density matrix $\rho(x,y;\tau)$ can be related to the initial one $\rho_0(x,y)$ by the expression
\begin{equation}
 \rho(x,y;\tau)=\frac1f\rho_0\left(\frac xf,\frac yf\right)\exp\left(-\frac if\frac{\dot{f}}{\omega_0}\frac{x^2-y^2}{2l_0^2}\right),
 \label{eq:obdm_scaling}
\end{equation}
where the scaling parameter is $f(\tau)=\sqrt{1+(\omega_0 \tau)^2}$ for the expansion in the absence of a trap and $l_0=\sqrt{\hbar/m\omega_0}$ is the initial trapping length in the continuum. The parameters $\omega_0$ and $m$ can be conveniently transformed into their lattice counterparts via $\hbar\omega_0=2\sqrt{V_2 t}$ and $m=\hbar^2/(2ta^2)$.

\begin{figure}[!t]
 \centering
 \includegraphics[width=0.49\linewidth]{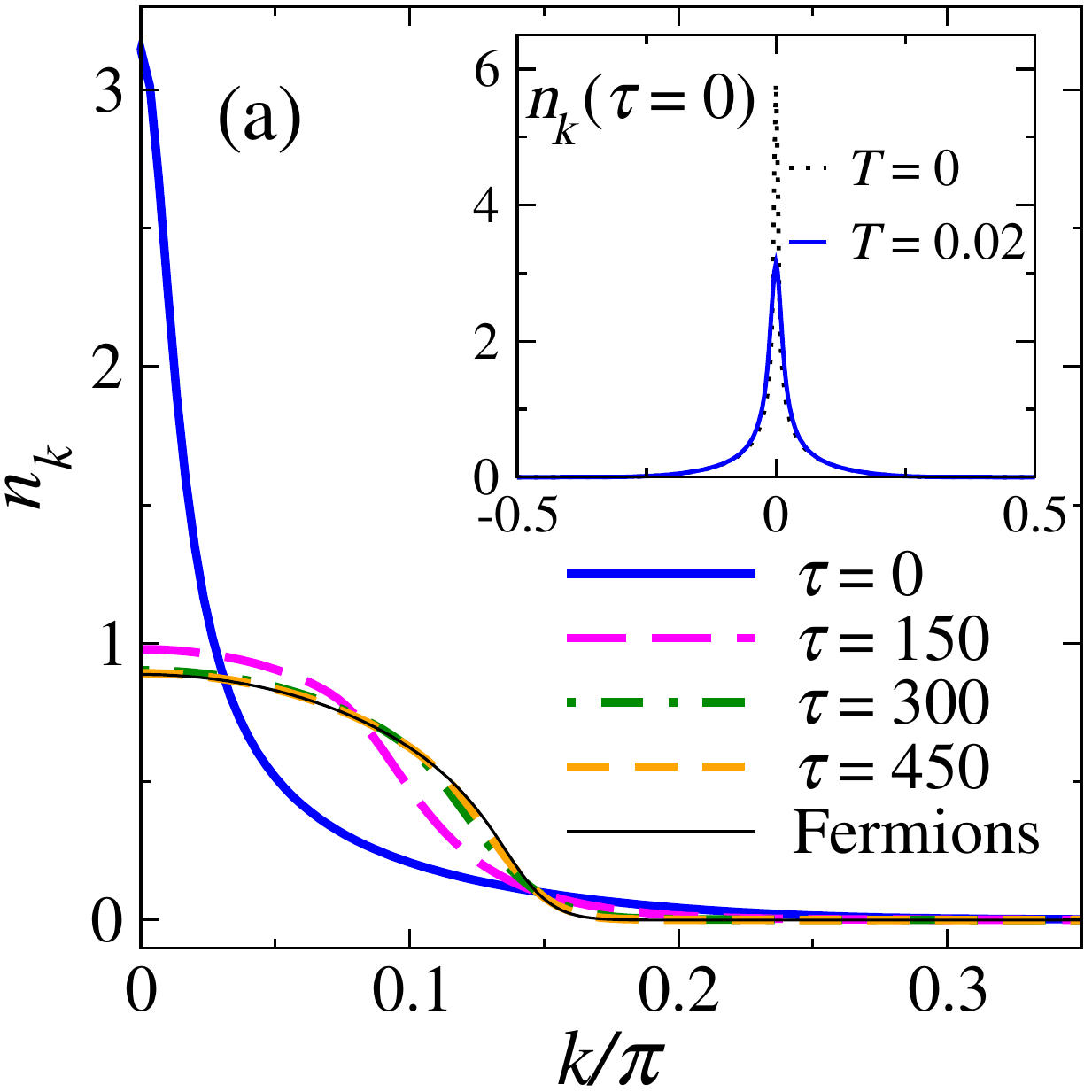}
 \includegraphics[width=0.495\linewidth]{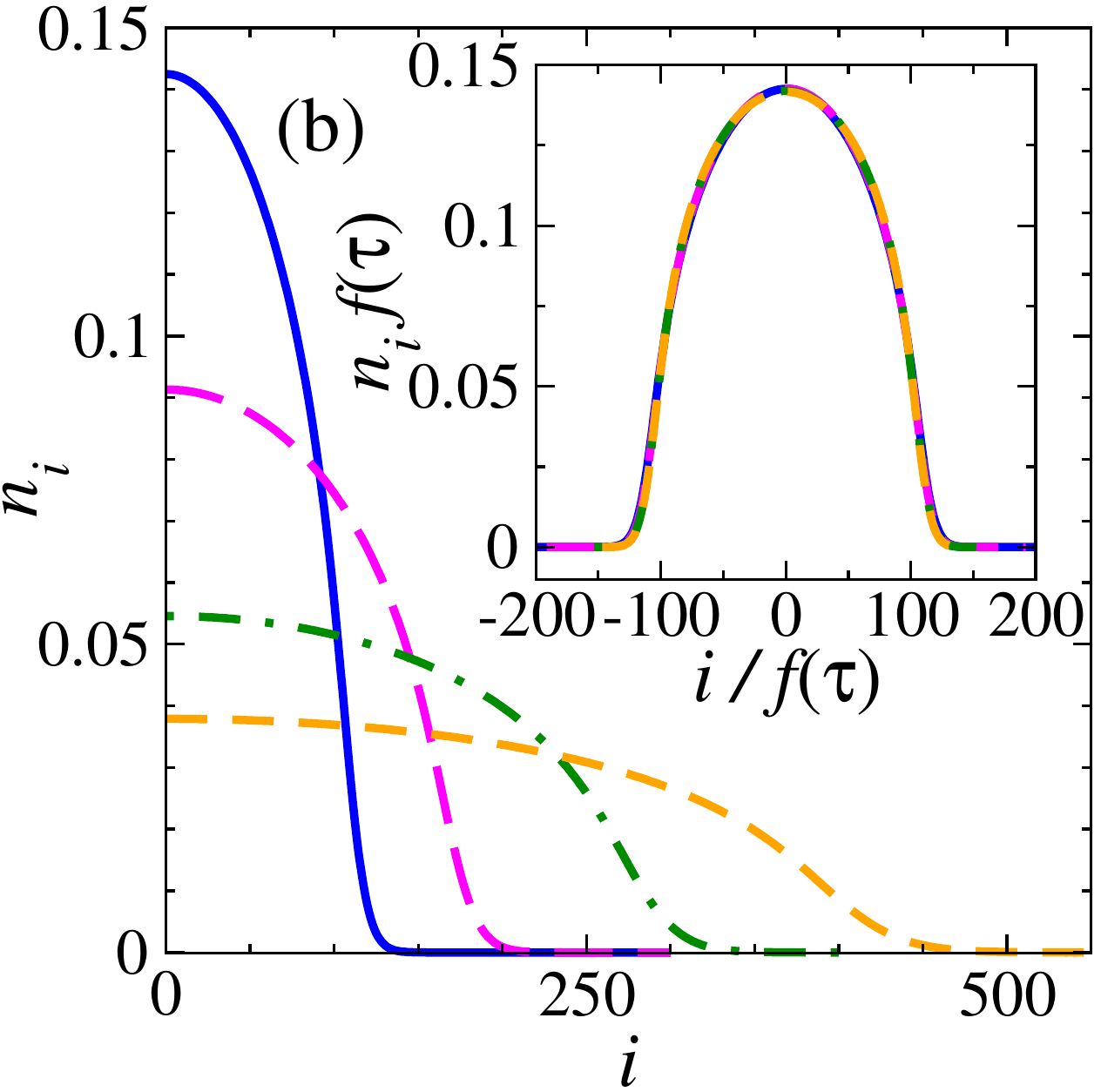}
 \caption{(Color online) (a) Momentum ($n_k$) and (b) site ($n_i$) occupations for $N=25$ hard-core bosons during the expansion in a lattice. The system is initially prepared in a harmonic trap with characteristic density $\tilde{\rho}=0.1$ and temperature $T=0.02$. The inset in panel (a) shows a comparison between $n_k$ at $\tau=0$ for $T=0.02$ and $T=0$, and highlights the effect of $T\neq0$. The inset in panel (b) makes apparent the self-similar behavior of the site occupations $n_i$ during the expansion.}
 \label{fig:fermionization}
\end{figure}

Figure~\ref{fig:fermionization} shows an example of the dynamics for an initial state with $\tilde{\rho}=0.1$ and $T=0.02$. During the early stages of the expansion [main panel in Fig.~\ref{fig:fermionization}(a)], the peak in $n_k$ at $k=0$ is rapidly suppressed. This is followed by a redistribution of $n_k$ (slower at high momenta) to fully match the corresponding momentum distribution of fermions. In contrast, the dynamics of the site occupations is self-similar, as expected from Eq.~\eqref{eq:obdm_scaling}, and as shown in the main panel of Fig.~\ref{fig:fermionization}(b) and its inset. The inset in Fig.~\ref{fig:fermionization}(a) shows the difference between the initial momentum distribution at $T=0.02$ and the ground state result. It highlights the effect of having a nonzero initial temperature in $n_k$.

\begin{figure}[!t]
 \centering
 \includegraphics[width=1.0\linewidth]{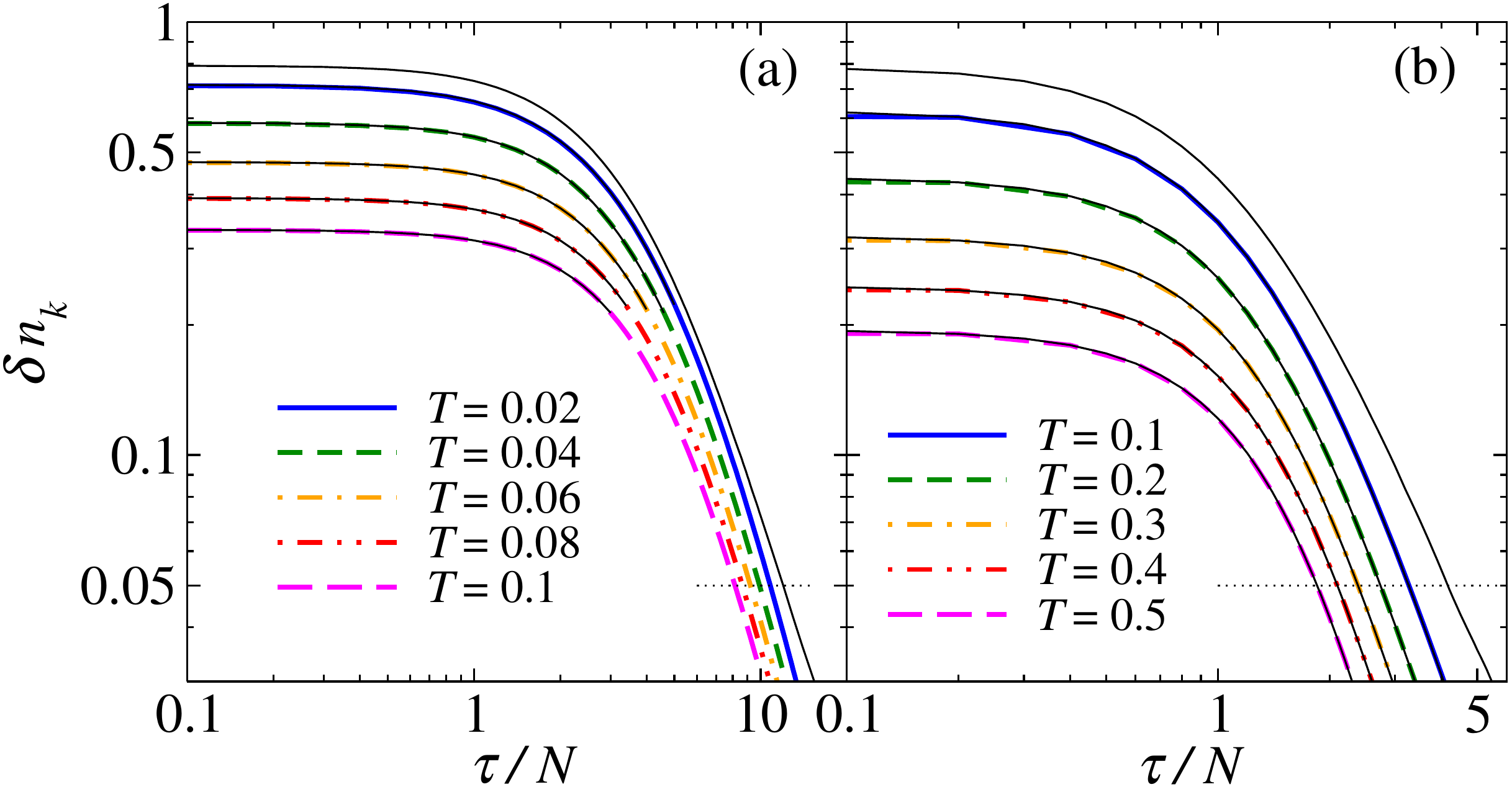}
 \caption{(Color online) Decrease of the relative difference $\delta n_k$ during the expansion for different initial characteristic densities: (a) $\tilde{\rho}=0.1$ and (b) $\tilde{\rho}=0.32$. Overlapping thick (color) and thin (black) lines show results for $N=25$ and $N=50$, respectively, and different values of $T$. The topmost black line depicts the ground state result for $N=50$. Horizontal dotted lines mark $\delta n_k=0.05$. The values of $\tau/N$ vs $T$ for $\delta n_k=0.05$ are reported in Fig.~\ref{fig:indicator}.}
 \label{fig:deltank}
\end{figure}

In order to demonstrate, quantitatively, that the momentum distribution of hard-core bosons indeed approaches that of noninteracting fermions, we compute the relative difference $\delta n_k(\tau)=\sum_k|n_k(\tau)-n_k^f|/\sum_k n_k^f$, where $n_k^f$ is the momentum distribution of the fermions~\cite{Rigol_Muramatsu_05b}. In Fig.~\ref{fig:deltank}, we report results for systems with $N=25$ and $N=50$ particles. In all cases, one can see that $\delta n_k(\tau)$ decreases with time. Note the data collapse for nonzero initial temperature when $\delta n_k(\tau)$ is plotted vs $\tau/N$, so we expect that the results will not change if the number of particles is further increased while keeping the characteristic density constant. It is also apparent in the figure that the results for $\tilde{\rho}=0.1$ [Fig.~\ref{fig:deltank}(a)] are qualitatively similar to those for $\tilde{\rho}=0.32$ [Fig.~\ref{fig:deltank}(b)], which means that the behavior observed is robust to changes in $\tilde{\rho}$, for small enough $\tilde{\rho}$.

\begin{figure}[!b]
 \centering
 \includegraphics[width=0.49\linewidth]{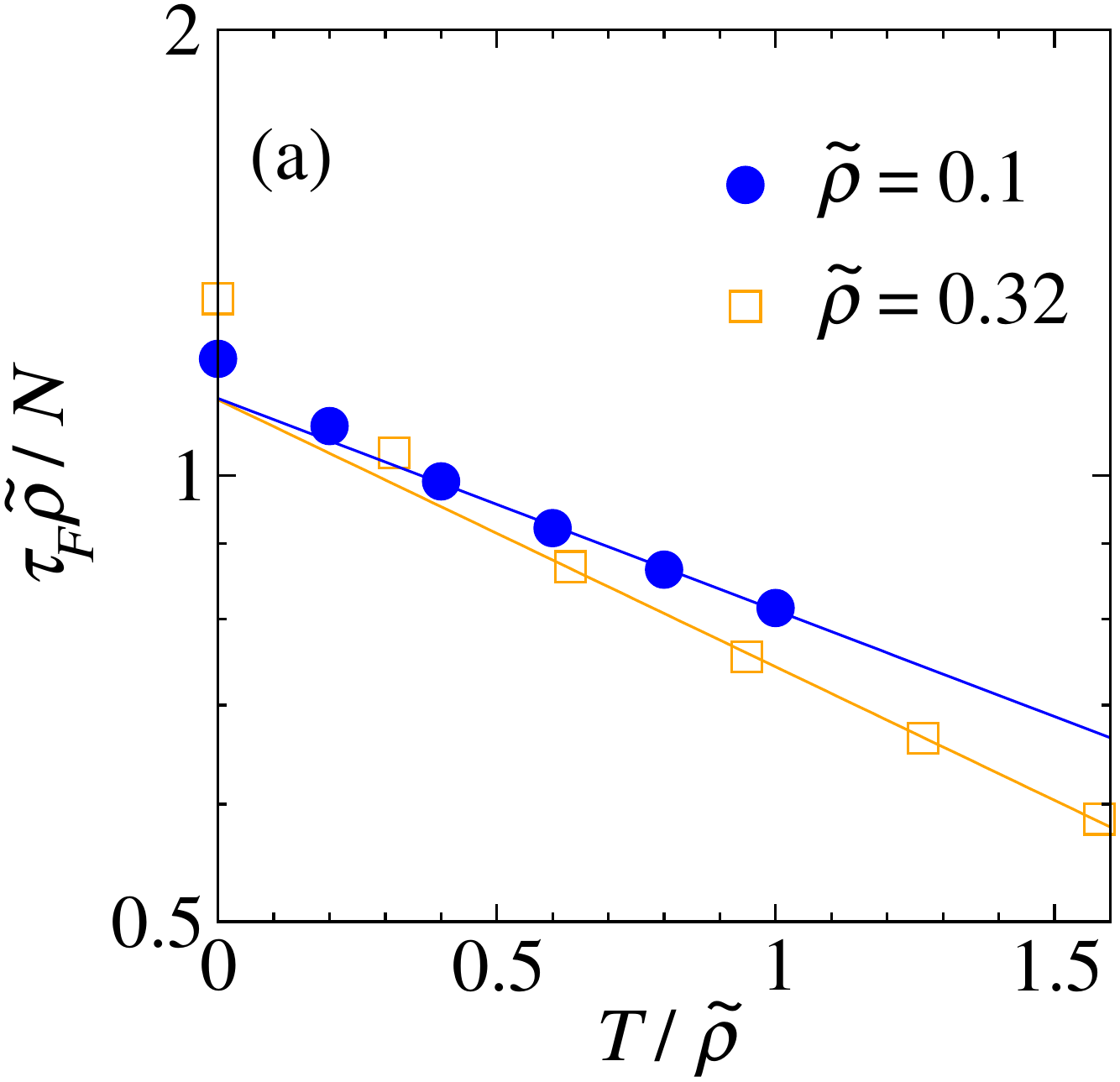}
 \includegraphics[width=0.49\linewidth]{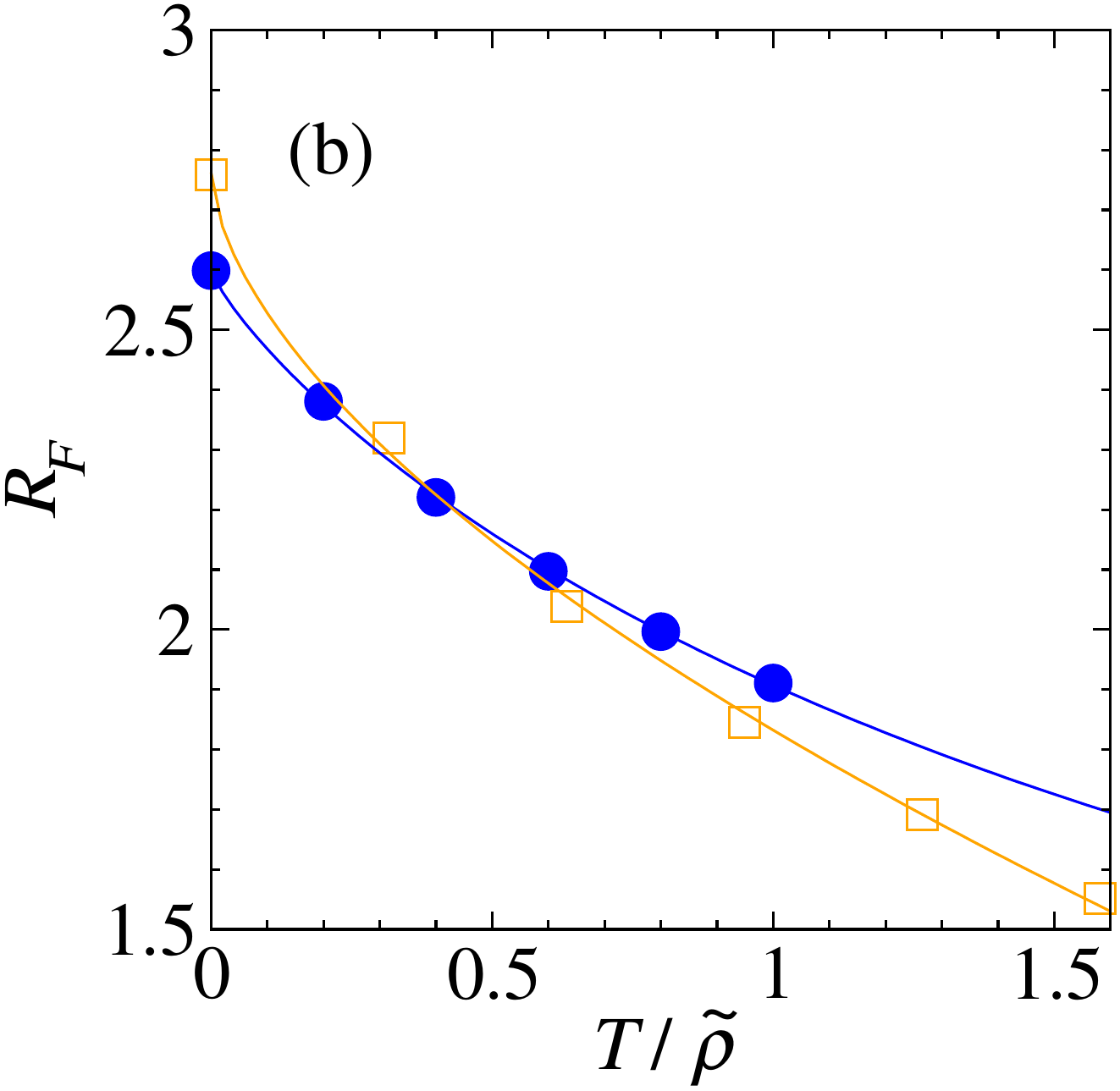}
 \caption{(Color online) (a) Time $\tau_F$ at which $\delta n_k=0.05$ (see Fig.~\ref{fig:deltank}) as a function of the effective temperature $T/\tilde{\rho}$. The solid lines are exponential fits to the numerical results. (b) Ratio $R_F$ between the full width at half maximum (FWHM) at $\tau_F$ and the FWHM at $\tau=0$ as a function of $T/\tilde{\rho}$. The solid lines are a guide to the eye.}
 \label{fig:indicator}
\end{figure}

Next, we analyze at which times the relative difference reaches a given small value, say $\delta n_k=0.05$ as marked by horizontal dotted lines in Fig.~\ref{fig:deltank}, as a function of the temperature of the initial state. The results of this analysis are reported in Fig.~\ref{fig:indicator}(a). They show that the rescaled time decreases, rather slowly but maybe exponentially, with increasing $T$. (The initial differences between the momentum distribution of hard-core bosons and fermions also decreases with increasing $T$.) Of more direct relevance to experiments is how the ratio between the cloud size at the time at which $\delta n_k=0.05$ and the initial cloud size ($R_F$) changes with increasing $T$. Results for this quantity, using the full width at half maximum as a measure of the cloud size, are presented in Fig.~\ref{fig:indicator}(b). That figure shows that $R_F$ decreases with increasing $T$, i.e., the fermionization of the momentum distribution of hard-core bosons may be easier to observe experimentally in systems that are initially at finite temperature.

\subsection{Natural orbitals and one-body correlations}

\begin{figure}[!b]
 \centering
 \includegraphics[width=1.0\linewidth]{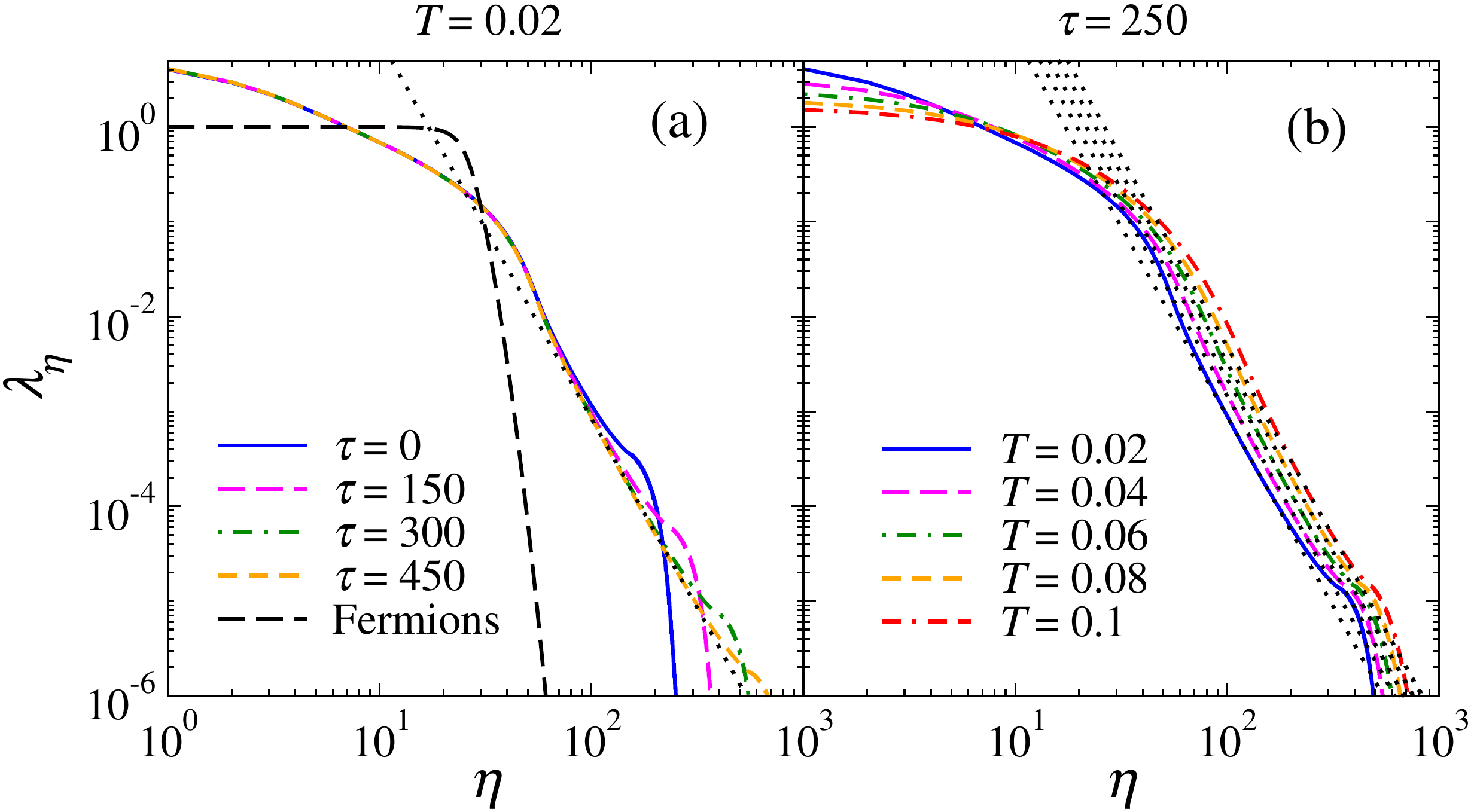}
 \caption{(Color online) (a) Natural orbital occupations for the same initial parameters and times for which momentum and site occupations are reported in Fig.~\ref{fig:fermionization}. The black dashed line depicts the natural orbital occupations of the noninteracting fermions, which do not change with time. (b) Natural orbital occupations at $\tau=250$ for different initial temperatures $T$. All black dotted lines are $\eta^{-4}$ fits to the tails.}
 \label{fig:natorb_occ}
\end{figure}

While the dynamical fermionization of the momentum distribution may lead one to conclude that the one-body correlations of hard-core bosons are approaching those of a system of noninteracting spinless fermions, this is not the case. This becomes clear if one studies the population of the natural orbitals, which are the eigenfunctions of the one-body density matrix~\cite{Penrose_56}
\begin{equation}
 \sum_j\rho_{ij}(\tau)\phi^\eta_j(\tau)=\lambda_\eta(\tau)\phi^\eta_i(\tau).
\end{equation}

In Fig.~\ref{fig:natorb_occ}(a), we show the natural orbital occupations at different times. It is remarkable that they almost do not change for small and intermediate values of $\eta$ (the lowest natural orbital occupation, $\lambda_0$, slightly increases), and that they are very different from those of a system of noninteracting fermions. All that happens during the expansion of hard-core bosons is that the tails $\lambda_\eta\propto\eta^{-4}$, known to occur in equilibrium \cite{Rigol_Muramatsu_04a}, extend to larger values of $\eta$. These changes can be attributed to the presence of the underlying lattice. [In the continuum limit ($\tilde{\rho}\rightarrow0$), $\lambda_\eta(\tau)$ is time independent.] Hence, the natural orbital occupations allow one to distinguish hard-core bosons from noninteracting fermions. Figure~\ref{fig:natorb_occ}(b) shows the natural orbital occupations for different initial temperatures after the same expansion time. As in the initial state, as $T$ increases, the population of the lowest natural orbitals decreases and the population of the tails increases, i.e., the prefactor of the $\eta^{-4}$ tails increases. This is similar to the behavior of the momentum distribution of hard-core bosons in thermal equilibrium in the continuum, for which the prefactor of the $k^{-4}$ momentum tails also increases with temperature \cite{Vignolo_13, xu_rigol_15}.

\begin{figure}[!t]
 \includegraphics[width=0.492\linewidth]{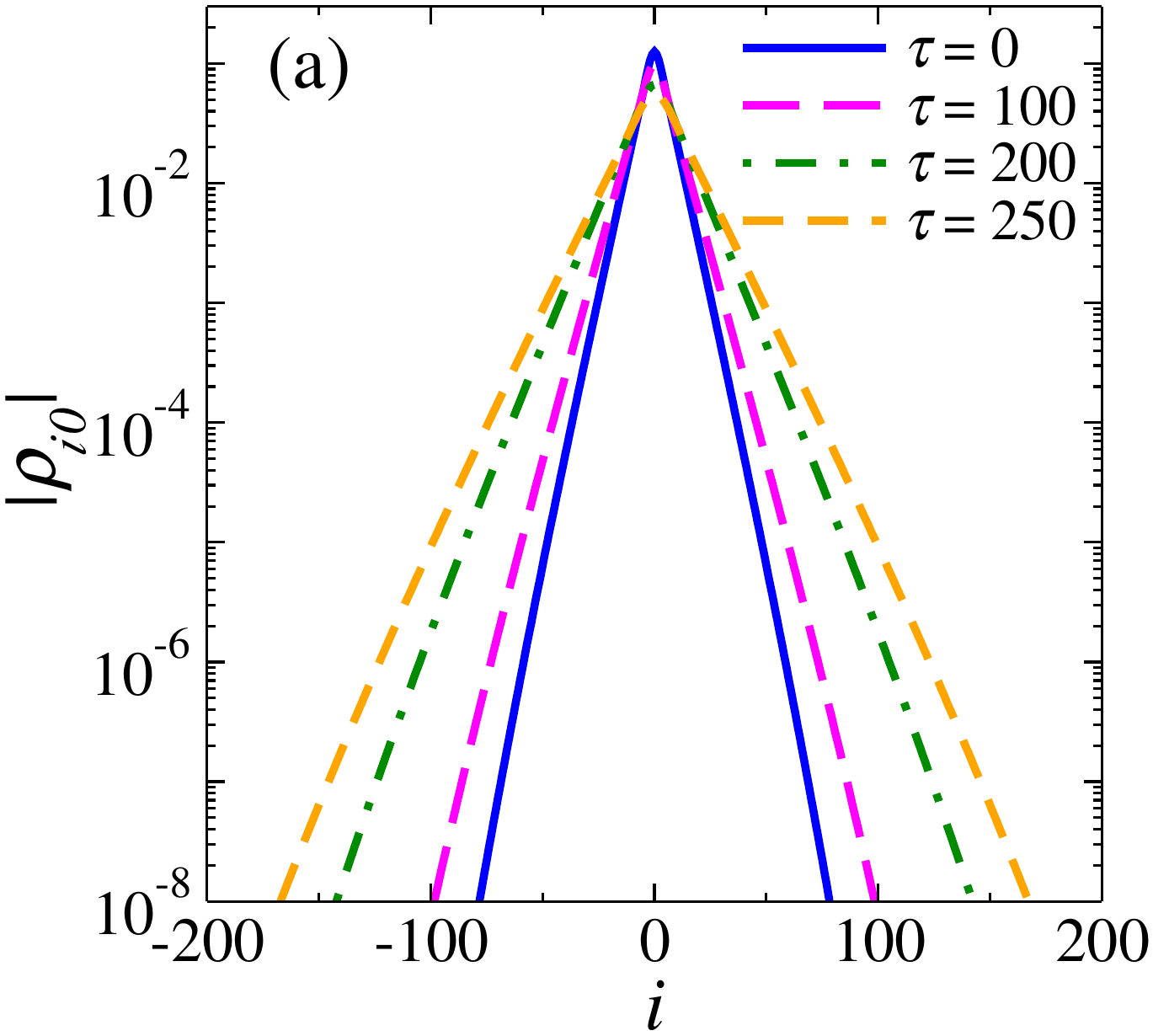}
 \includegraphics[width=0.493\linewidth]{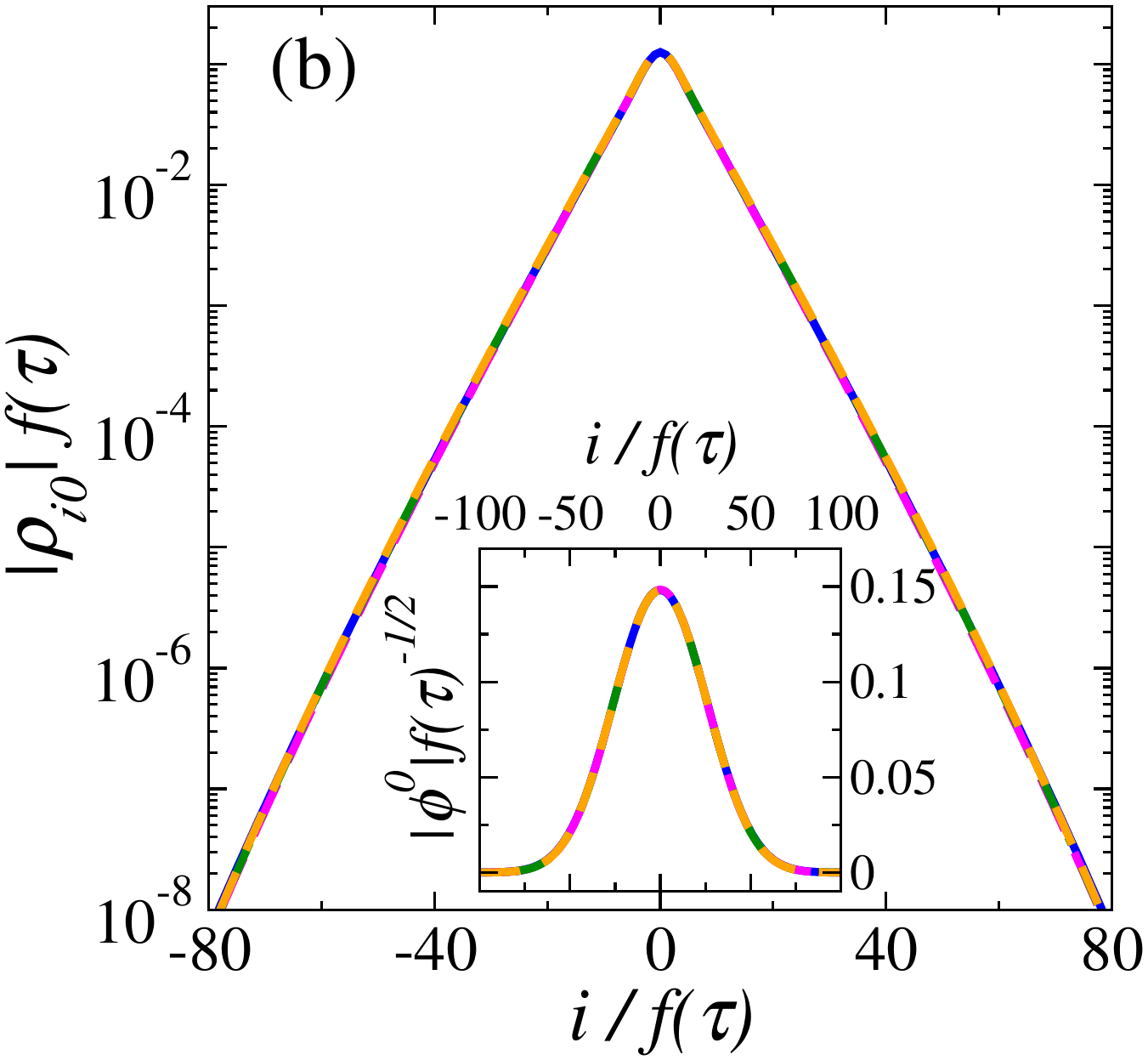}
 \caption{(Color online) (a) One-body density matrix at different times for $N=25$, $\tilde{\rho}=0.1$, and $T=0.1$. (b) Scaled results for the one-body density matrix (main panel), and the lowest natural orbital wavefunction (inset).}
 \label{fig:obdm_scaling}
\end{figure}

The fact that the natural orbitals remain (mostly) unchanged during the expansion follows from the scaling solution of one-body correlations in Eq.~\eqref{eq:obdm_scaling}. In Fig.~\ref{fig:obdm_scaling}(a), we show how one-body correlations decay away from the center of our lattice system at different times. Figure~\ref{fig:obdm_scaling}(b) shows that both, the one-body correlations as well as the lowest natural orbital, are well described by the scaling solution.

\section{Quasicondensation}\label{Sec:IV}

\subsection{Dynamics from a linear potential}

For the symmetric expansion starting from an initial pure Fock state with one particle per site, it was shown in Ref.~\cite{Rigol_Muramatsu_04b} that one-body quasi-long-range correlations develop dynamically and lead to the emergence of quasi-condensates at finite momentum $k=\pm\pi/2$. Our goal in this section is to understand what happens when the initial state is not a pure state but a mixed thermal equilibrium state. During the melting of a Mott insulator, while there are sites in the center with occupation one, the symmetric expansion can be described as two independent expansions to the left and to the right. 

Here we focus on the expansion of finite-temperature Mott domains in one direction after turning off the confining potential. We construct them as thermal equilibrium states in the presence of a linear potential ($\alpha=1$). For this potential, which has been studied in multiple works in the literature~\cite{Eisler_09, lancaster10, lancaster16, Vidmar_Iyer_15}, we obtain an analytic understanding of the effective cooling observed during the free expansion dynamics. We note that these systems are not parity symmetric as the ones considered in the previous section in the harmonic potential.

\begin{figure}[!t]
 \centering
 \includegraphics[width=1.0\linewidth]{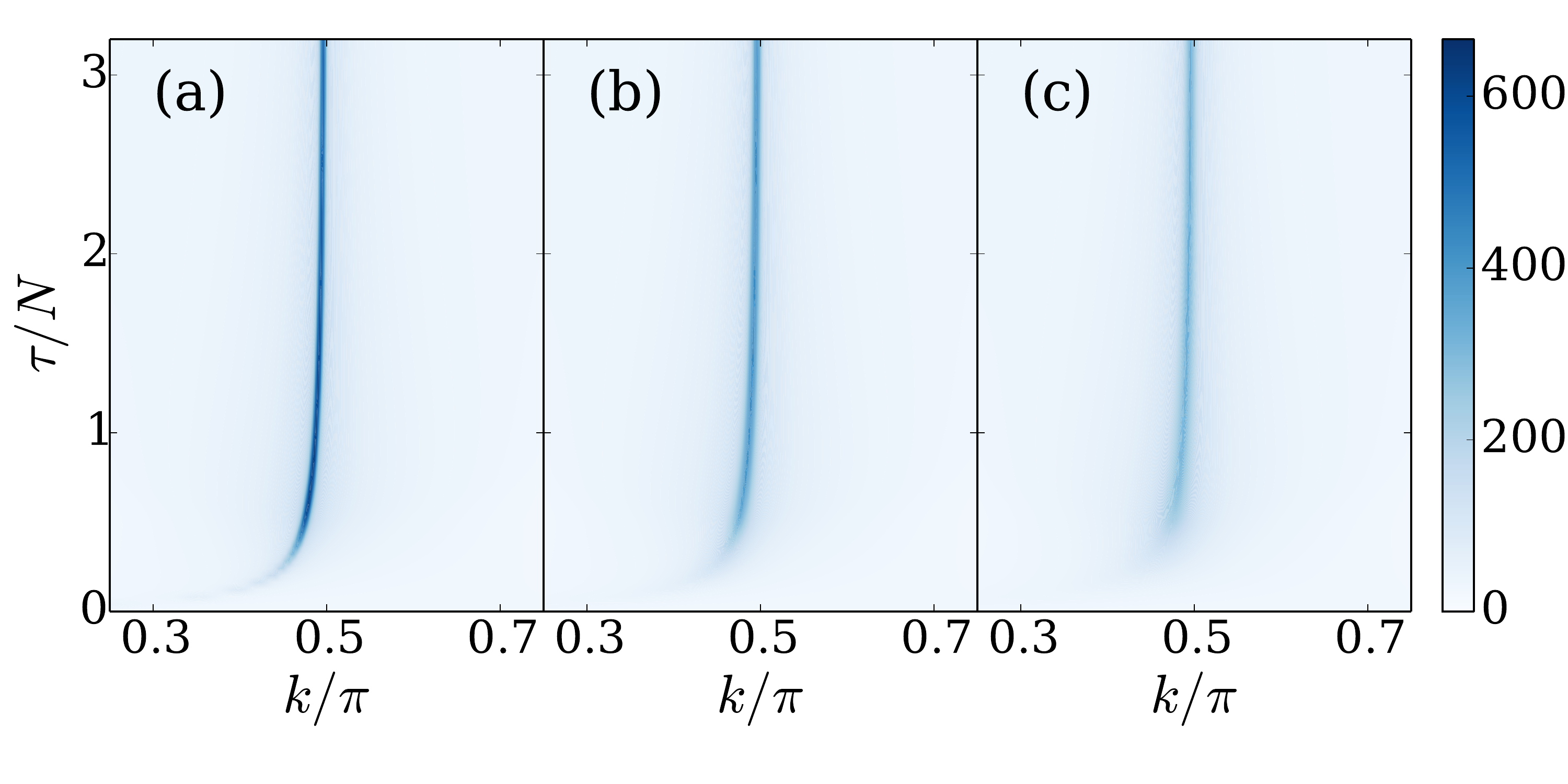}
 \caption{(Color online) Contour plots of the momentum distribution as a function of the expansion time. Panels (a)-(c) correspond to initial temperatures $T=0.1$, $T=0.5$, and $T=1.0$, respectively. These systems contain $N=200$ particles and, initially, $V_1=0.125$. Open boundaries are considered to, after the linear potential is turned off, constrain the expansion of the Mott domain to the right.}
 \label{fig:contour_perfilK}
\end{figure}

In Fig.~\ref{fig:contour_perfilK}, we show (color) contour plots of the momentum distribution as a function of time for three initial temperatures of the Mott domains. For all temperatures, one can see that a peak emerges in the momentum distribution. The height of the peak increases with time and its position approaches $k=\pi/2$ \cite{Vidmar_Iyer_15}. As the temperature increases, from Fig.~\ref{fig:contour_perfilK}(a) to~\ref{fig:contour_perfilK}(c), the peak becomes wider and its height decreases. The emergence of the peak during the expansion indicates that the system develops off-diagonal one-body correlations. They appear to weaken as the initial temperature is increased.

During the melting of a perfect Mott insulator with one particle per site, which is the $T=0$ and $V_1=\infty$ limit of our setup, the highest occupied momentum mode and the lowest natural orbital (the highest occupied one) exhibit occupations $n_k^m(\tau)\propto\sqrt{\tau^3}$ and $\lambda_0(\tau)\propto\sqrt{\tau}$~\cite{Rigol_Muramatsu_04b, Rigol_Muramatsu_05c}, respectively. They reach a maximum shortly after no sites with occupation one remain in the system, and then start to decrease slowly. The maximum values reached depend on the number of particles: $[n_k^m]_\text{max}\propto\sqrt{N}$ and $[\lambda_0]_\text{max}\propto\sqrt{N}$ and occur at a time that increases linearly with $N$~\cite{Rigol_Muramatsu_04b, Rigol_Muramatsu_05c}. Furthermore, there exists scaling solutions to $n_k^m(\tau)/\sqrt{N}$ and $\lambda_0(\tau)/\sqrt{N}$ as a function of $\tau/N$. Such a universal behavior breaks down at nonzero temperature. However, in the early times of the expansion, $n_k^m(\tau)\approx\text{const}\,\sqrt{\tau^3}$ and $\lambda_0(\tau)\approx\text{const}\,\sqrt{\tau}$.

\begin{figure}[!t]
 \centering
 \includegraphics[width=0.525\linewidth]{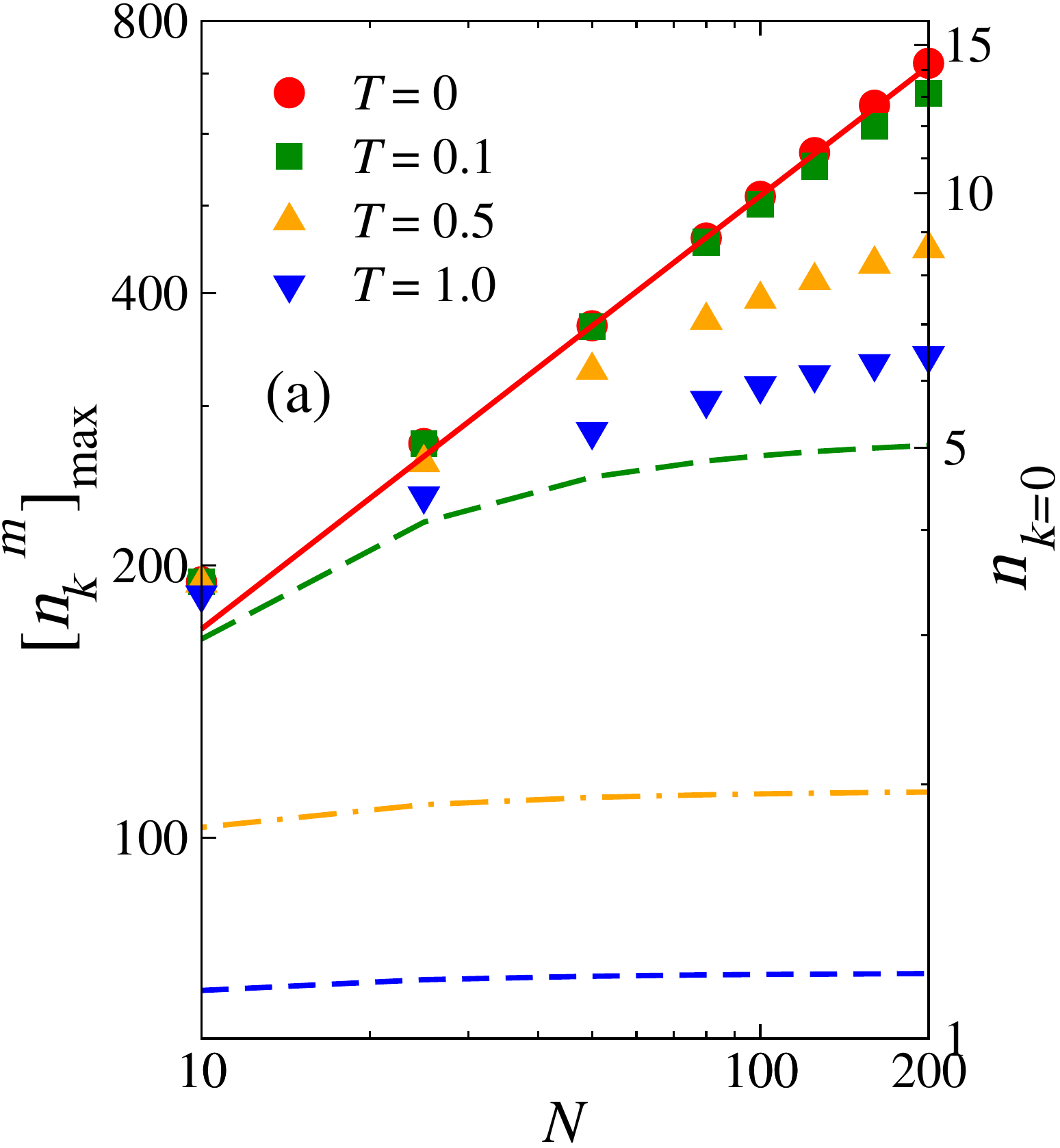}\hspace{0.025cm}
 \includegraphics[width=0.46\linewidth]{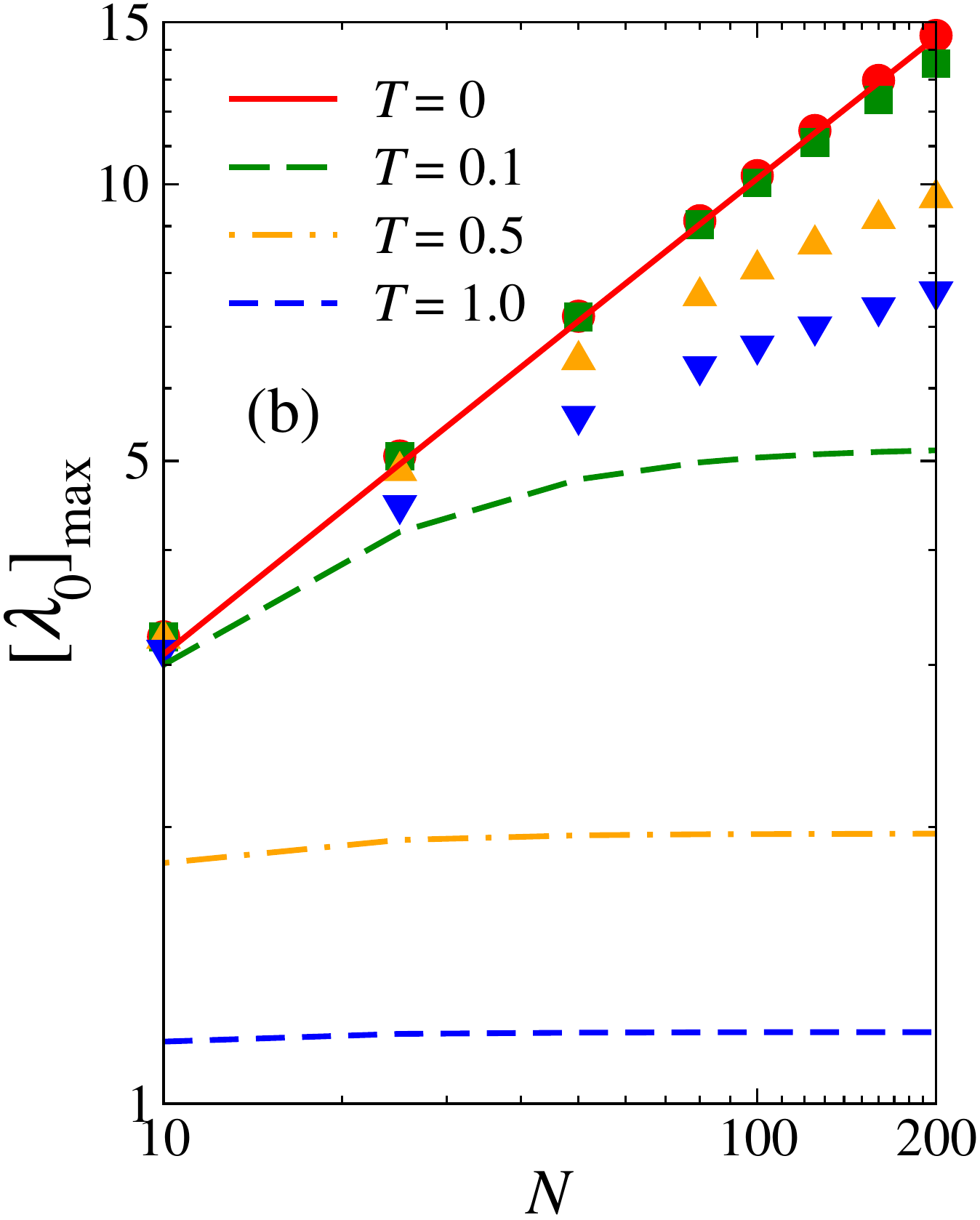}
 \caption{(Color online) Maximal values (a) $[n_k^m]_\text{max}$ and (b) $[\lambda_0]_\text{max}$ during the expansion (symbols) as a function of the number of bosons in the system. The calculations are done for initial states with constant $\tilde{\rho}=50$, and different $T=0.1$, 0.5 and 1.0. We also report the occupation of: (a) the zero-momentum mode $n_{k=0}$ (labels on the right), and (b) the lowest natural orbital $\lambda_0$, in homogeneous lattices with open boundary conditions at half-filling (lines) at the same initial temperatures of the systems that undergo expansion.}
 \label{fig:saturation_peaks}
\end{figure}

In Fig.~\ref{fig:saturation_peaks}, we plot $[n_k^m]_\text{max}$ [Fig.~\ref{fig:saturation_peaks}(a)] and $[\lambda_0]_\text{max}$ [Fig.~\ref{fig:saturation_peaks}(b)] vs $N$ for different initial temperatures. As in the previous section, when $N$ is increased $V_1$ is decreased so that $\tilde{\rho}=NV_1$ remains constant in the initial state. At $T=0$, $[n_k^m]_\text{max}$ and $[\lambda_0]_\text{max}$ display a power-law increase with $N$ at large $N$ (with exponent 1/2). For nonzero initial temperatures, $[n_k^m]_\text{max}$ and $[\lambda_0]_\text{max}$ still can be seen to increase with $N$. One can wonder whether this growth is consistent with the growth of $n_{k=0}$ and $\lambda_0$ in the initial thermal equilibrium state. In Fig.~\ref{fig:saturation_peaks}, we also report results for the latter quantities in a box trap at half-filling at the initial temperatures of the systems that undergo the expansion. For $T=0$, the equilibrium results closely follow those of the dynamics for large $N$. However, for nonzero temperature, both $n_{k=0}$ and $\lambda_0$ in equilibrium exhibit a much slower growth and saturate for large $N$. This indicates that if an effective thermal equilibrium description of the expansion is possible, it will likely involve lower temperatures than those of the initial state.

\subsection{Emergent eigenstate solution}

In the thermodynamic limit, an exact description of the dynamics discussed in the previous subsection can be obtained using the emergent local Hamiltonian introduced in Ref.~\cite{Vidmar_Iyer_15}. There it was shown that the time-evolving state generated by the expansion of the ground state of Hamiltonian \eqref{eq:ham}, with $\alpha=1$ (linear potential), is the ground state of the following emergent Hamiltonian
\begin{equation}
{\cal \hat H}(\tau)  =  -{\cal A}(\tau) \sum_l (e^{i \varphi(\tau)} \hat f_{l+1}^\dagger \hat f^{}_l + {\rm H.c.})  + \, V_1 \sum_l l \, \hat n_l ,  \label{eq:eHam}
\end{equation}
where the effective hopping amplitude is
\begin{equation}\label{eq:ehopp}
{\cal A}(\tau) = \sqrt{1 + (V_1 \tau)^2},
\end{equation}
and the time-dependent phase is
\begin{equation} \label{Eq:ephase}
\varphi(\tau) = \arctan{\left({V_1 \tau}\right)}.
\end{equation}
The same applies to excited states, so long as they contain a region with site occupation one in the left edge of the system and zero in the right one. Hence, the emergent local Hamiltonian can be used to describe finite-temperature initial states if the temperature is not too high. The regime of validity extends to higher temperatures with increasing the initial value of $V_1$.

\begin{figure}[!t]
 \centering
 \includegraphics[width=0.48\linewidth]{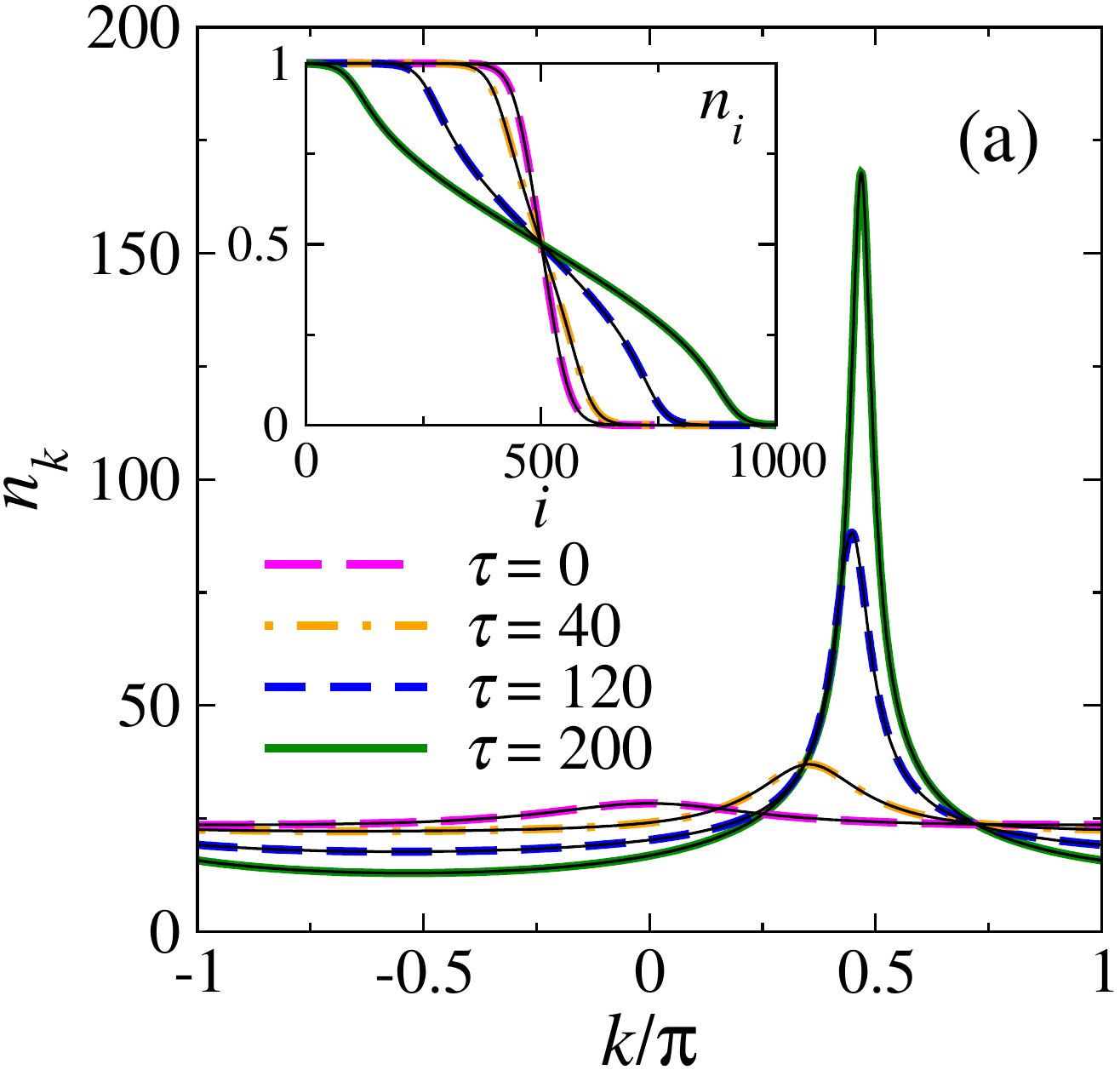}
 \includegraphics[width=0.499\linewidth]{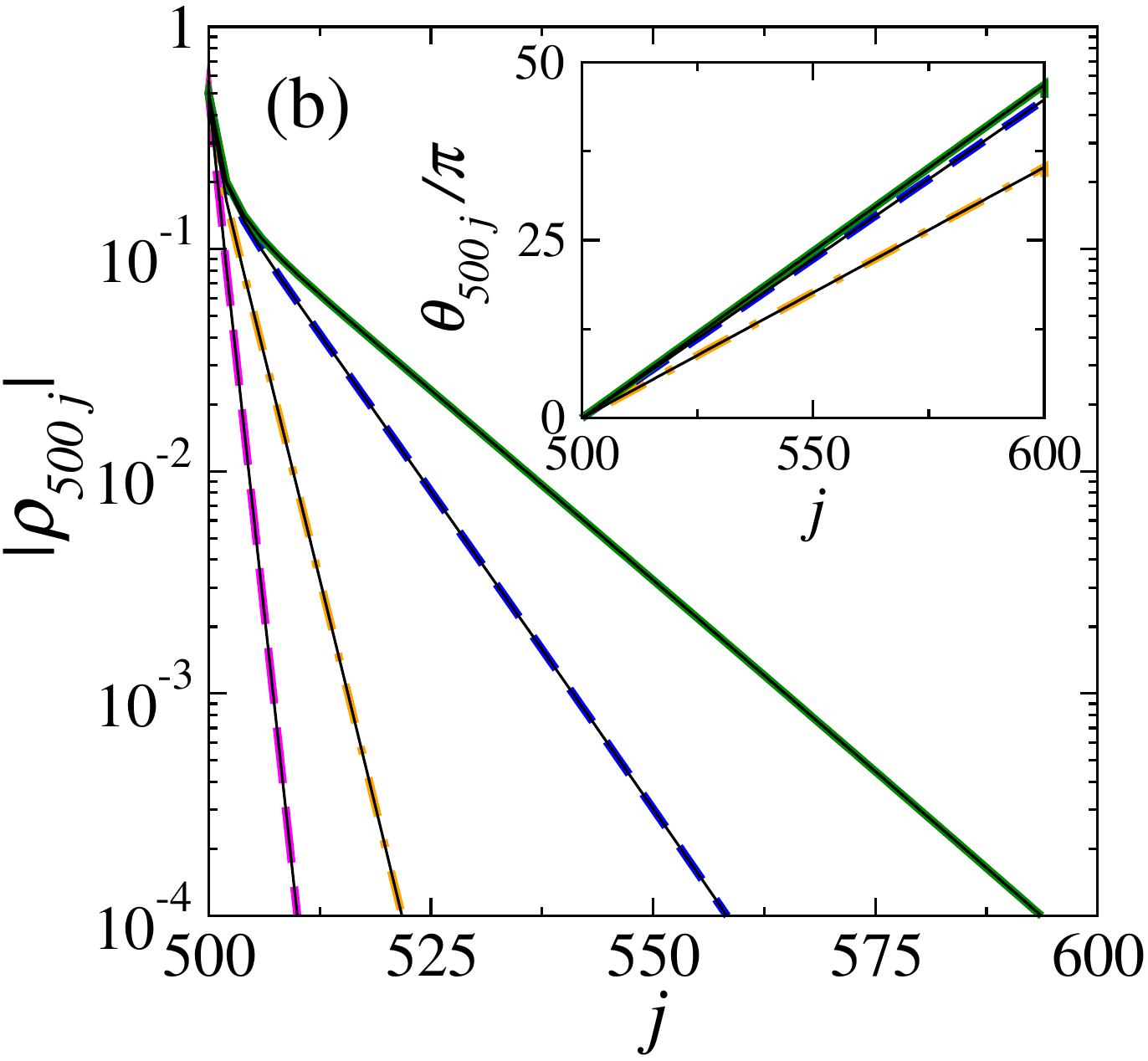}
 \caption{(Color online) Comparison between the emergent local Hamiltonian description [lines in the legend in panel (a)] and the exact results from the dynamics (thin continuous lines). The calculations were done in a system with $N=500$, an initial potential with $V_1=0.05$, and $T=1.0$, and results are reported for four times. (a) Momentum distributions (main panel) and site occupation profiles (inset). (b) Absolute values of one-body correlations measured from site 500 (main panel) and the corresponding phases (inset).}
 \label{fig:densm_cooling}
\end{figure}

The description in terms of the emergent Hamiltonian \eqref{eq:eHam} makes apparent why an effective cooling is taking place in the system. The initial temperature sets which eigenstates of the initial Hamiltonian are involved in the dynamics. Hence, $T$ does not depend on $\tau$ in the emergent local Hamiltonian description. On the other hand, the effective hopping amplitude ${\cal A}(\tau)$ increases with time. This results in a ratio between $T$ and ${\cal A}(\tau)$ that decreases with time. One can think of the time evolving state as a thermal equilibrium state with temperature
\begin{equation}
 T_\text{eff}(\tau)=T/\sqrt{1+(V_1\tau)^2} \;.
 \label{eq:T_eff}
\end{equation}
in the following effective Hamiltonian:
\begin{equation}
{\cal \hat H}_{\text{eff}}(\tau)  =  -\sum_l (e^{i \varphi(\tau)} \hat f_{l+1}^\dagger \hat f^{}_l + {\rm H.c.})  + \, V_1(\tau) \sum_l l \, \hat n_l ,  \label{eq:eeHam}
\end{equation}
where $V_1(\tau)=V_1/{\cal A}(\tau)=V_1/\sqrt{1 + (V_1 \tau)^2}$. This means that if at time $\tau$ after the initial linear potential has been turned off one suddenly quenches the Hamiltonian from the free one, $\hat H_0=-t\sum_l(\hat b_{l+1}^\dag \hat b_l+\text{H.c.})$, dictating the dynamics to ${\cal \hat H}_{\text{eff}}(\tau)$ all evolution will stop as the system will be in a thermal equilibrium state of ${\cal \hat H}_{\text{eff}}(\tau)$ with temperature $T_\text{eff}(\tau)$. This is something that could potentially be done in experiments with ultracold quantum gases to test the validity of the emergent Hamiltonian description.

In Fig.~\ref{fig:densm_cooling}(a), we compare the exact results obtained from the quantum dynamics for the momentum distribution (main panel) and the site occupations (inset) with those obtained using the emergent local Hamiltonian description from Eqs.~\eqref{eq:T_eff} and \eqref{eq:eeHam}. The results from both calculations are indistinguishable from each other. In the main panel in Fig.~\ref{fig:densm_cooling}(b), we compare the results from the exact dynamics and the effective equilibrium description for the absolute value of one-body correlations. They are also identical, and exhibit an exponential decay with a correlation length that increases with $\tau$, as expected from the emergent local Hamiltonian description. The latter also allows one to understand why the position of the momentum peak moves toward larger values of $k$. This is the result of the change of the phase \eqref{Eq:ephase} with $\tau$ \cite{Vidmar_Iyer_15}. The inset in Fig.~\ref{fig:densm_cooling}(b) shows the phase, from $\rho_{lj}=|\rho_{lj}|e^{i\theta_{lj}}$ for $l=500$, corresponding to the absolute values in the main panel. One case see that $\theta_{lj}$ increases linearly with the distance $l-j$ and that the slope increases with $\tau$.

As mentioned before, the effective equilibrium description based on Eqs.~\eqref{eq:T_eff} and \eqref{eq:eeHam} applies only while there are sites with occupation one in the left boundary of the chain and zero in the right one \cite{Vidmar_Iyer_15}. Thus the minimum effective temperature is determined by Eq.~\eqref{eq:T_eff} at the time $\tau_c$ at which the occupation in the leftmost (rightmost) site departs from one (zero). If we assume that $L>2N$, the occupation of the leftmost site will depart from one before the occupation of the rightmost one departs from zero. Because of the Lieb-Robinson bound, the time for the former to occur increases linearly with $N$. Since the appropriate thermodynamic limit requires $\tilde{\rho}=V_1 N=$const, the minimum effective temperature is given by $T_\text{eff}^\text{min}=T/\sqrt{1+(\tilde{\rho}\tau_c/N)^2}$. Thus $T_\text{eff}^\text{min}$ is size independent when $N$ is large enough. The behavior of $[\lambda_0]_\text{max}$ and $[n_k^m]_\text{max}$ depicted in Fig.~\ref{fig:saturation_peaks} can be reproduced with the effective thermal equilibrium state with $T_\text{eff}^\text{min}$.

\subsection{Reference Hamiltonian}

Another, maybe more intuitive, way to understand the behavior observed during the expansion of the Mott domains of hard-core bosons is to think of the dynamical state as a thermal equilibrium state in a boosted reference frame, an idea that was explored in Ref.~\cite{Meisner_08} in the context of the expansion of ground-state Mott insulators in the Fermi-Hubbard model. In this picture, the effective cooling in the dynamical system can be understood to be the result of internal energy being converted into center-of-mass energy, leading to a continuous reduction of the internal energy in the reference frame. 

In order to construct the reference Hamiltonian $\hat H_\text{ref}$ in the comoving frame, one needs to modify the Hamiltonian dictating the dynamics by introducing a site-dependent potential of strength $\epsilon_l(\tau)$ to reproduce exactly the site occupations at time $\tau$. For the expansion from the ground state, such a reference Hamiltonian was constructed in Ref.~\cite{Eisler_09}
\begin{equation}
 \hat H_\text{ref}=-t\sum_l(\hat f^\dag_{l+1}\hat f_l+\text{H.c.})+V_1(\tau)\sum_l l\,\hat n_l\;.
 \label{eq:H_ref}
\end{equation}
It is nothing but the Hamiltonian in Eq.~\eqref{eq:eeHam} without the time-dependent phases in the hopping terms. Those phase factors do not alter the properties of observables such as the site occupations. 

\begin{figure}[!t]
 \centering
 \includegraphics[width=1.0\linewidth]{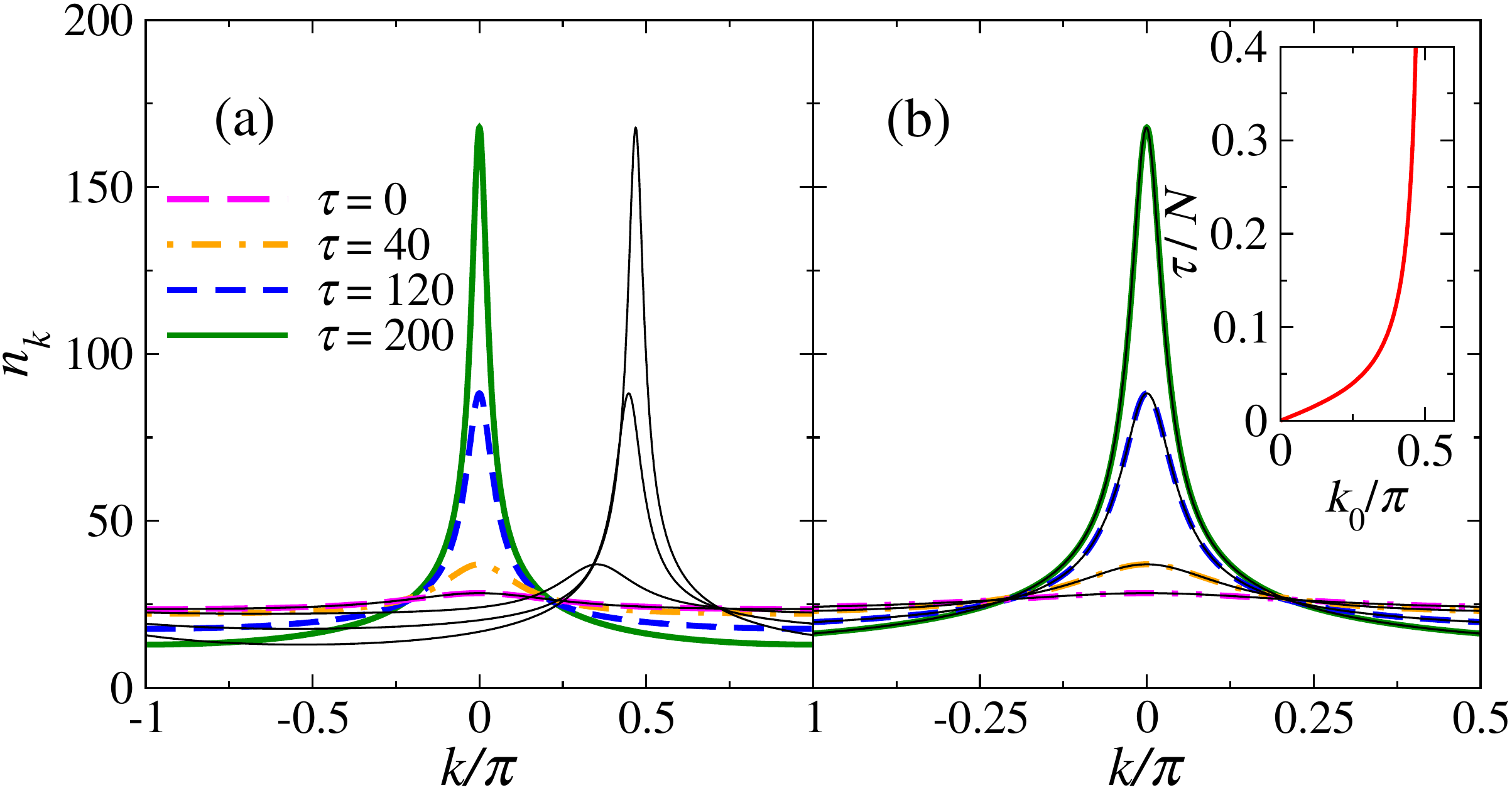}
 \caption{(a) Momentum distributions in the reference system [lines in the legend in panel (a)] compared to those during the expansion [thin lines, see also Fig.~\ref{fig:densm_cooling}(a)]. (b) Results after shifting $n_k$ during the expansion by the peak momentum $k_0$. (Inset) Peak momentum $k_0$ as a function of time. The calculations were done in a system with $N=500$, an initial potential with $V_1=0.05$, and $T=1.0$, and results are reported for four times (see also Fig.~\ref{fig:densm_cooling}).}
 \label{fig:nk_ref}
\end{figure}

For initial states at finite temperature, the effective temperature in the reference system is identical to the time-dependent effective temperature in the context of the emergent local Hamiltonian; see Eq.~\eqref{eq:T_eff}. In Fig.~\ref{fig:nk_ref}(a), we show the momentum distribution in the reference frame and in the expanding system. The peaks can be seen to be shifted by a time-dependent momentum. In the reference frame, the momentum distribution is symmetric about $k=0$. In Fig.~\ref{fig:nk_ref}(b), we show the results in the reference frame and in the expanding system after shifting the momentum distribution of the latter by the momentum $k_0(\tau)$ of the maximum of $n_k$, which is determined by the phase in Eq.~\eqref{eq:eeHam}: $k_0(\tau)=\varphi(\tau) = \arctan{\left({V_1 \tau}\right)}$ \cite{Vidmar_Iyer_15}. The distributions are now indistinguishable from each other. The dependence of $k_0$ on $\tau$ for this particular setup is shown in the inset in Fig.~\ref{fig:nk_ref}(b).

\section{Conclusions}\label{Sec:V}

We studied the sudden expansion of hard-core bosons in thermal equilibrium in 1D lattices. For low initial site occupations, we showed that the expansion is self-similar and that the momentum distribution fermionizes at long times. This had been shown to occur during the expansion from initial ground states in Refs.~\cite{Rigol_Muramatsu_05b, Minguzzi_05}. In addition, we showed that the natural orbital occupations almost do not change in time (they exhibit the characteristic $\eta^{-4}$ behavior of systems in thermal equilibrium) and are distinctively different from those of noninteracting fermions. This means that the natural orbital occupations can be used to distinguish hard-core bosons from fermions even after the momentum distribution of the former has fermionized.

For the expansion from an initial Mott insulator, we showed that the enhancement of one-body correlations known from studies of pure states \cite{Rigol_Muramatsu_04b} is robust against nonzero initial temperatures. However, increasing the temperature does weaken those correlations, and results in smaller peaks in the momentum distribution. Remarkably, the expansion leads to an effective cooling, namely, the system can be described by effective thermal equilibrium states with a correlation length that increases with time. We discussed two related ways to understand this phenomenon, one in terms of an emergent local Hamiltonian and the second one in terms of a thermal equilibrium state in a boosted reference frame. Our results explain why experiments with ultracold gases, such as the ones in Ref.~\cite{Vidmar_Ronzheimer_15}, should be able to observe large momentum peaks in $n_k$ even if the initial states are not in the ground state. 

\begin{acknowledgements}
This work was supported by the U.S. Office of Naval Research, Award No. N00014-14-1-0540. The computations were done in the Institute for CyberScience at Penn State and the Center for High-Performance Computing at the University of Southern California. We thank L. Vidmar and D. S. Weiss for insightful discussions.
\end{acknowledgements}

\bibliography{FreeExpansion_HCB}
\end{document}